\documentclass{fundam}

\usepackage{amsmath, amssymb, graphicx, color, rcs, qtree, moreverb}
\usepackage[mathscr]{euscript}
\usepackage[all]{xy}
\usepackage{caption}
\usepackage{array} 					
\usepackage{algorithm}				
\usepackage{algpseudocode}   

\definecolor{Green}{RGB}{0,128,0}
\newcommand{\red}[1]{\textcolor{red}{#1}}
\newcolumntype{L}[1]{>{\raggedright\let\newline\\\arraybackslash\hspace{0pt}}m{#1}}
\newcolumntype{C}[1]{>{\centering\let\newline\\\arraybackslash\hspace{0pt}}m{#1}}
\newcolumntype{R}[1]{>{\raggedleft\let\newline\\\arraybackslash\hspace{0pt}}m{#1}}

\def\str#1{\mbox{\boldmath $#1$}}

\def\itbf#1{\textit{\textbf{#1}}}
\def\s#1{\mbox{\boldmath $#1$}}

\newcommand{\SA}{\mathcal{SA}}
\newcommand{\LCP}{\mathcal{LCP}}
\newcommand{\RSF}{\mathcal{RSF}}
\newcommand{\OLP}{\mathcal{OLP}}
\newcommand{\ST}{\mathcal{ST}}

\newcommand{\df}[1]{\textit{\textbf{#1}}}
\newcommand{\bigO}{\mathcal{O}}

\begin{document}


\setcounter{page}{17}
\publyear{22}
\papernumber{2164}
\volume{190}
\issue{1}

\finalVersionForARXIV


\title{String Covering: A Survey}

\author{Neerja Mhaskar\thanks{Address for correspondence: Department of Computing and Software,
                     McMaster University, 1280 Main Street West, Hamilton, ON L8S 4L8, Canada.}
\\
Department of Computing and Software \\
McMaster University \\
1280 Main Street West, Hamilton, ON L8S 4L8, Canada\\
pophlin@mcmaster.ca
\and W.F. Smyth\thanks{This research was funded by NSERC [Grant No. 10536797]. \newline \newline
                    \vspace*{-6mm}{\scriptsize{Received Novemver 2022; \ accepted September 2023.}}  }
\\
Department of Computing and Software \\
McMaster University \\
1280 Main Street West, Hamilton,  ON L8S 4L8, Canada\\
smyth@mcmaster.ca
}

\maketitle

\runninghead{N. Mhaskar and W.F. Smyth}{String Covering: A Survey}

\vspace{3mm}
\begin{abstract}
  The study of strings is an important combinatorial field that precedes the digital computer.  Strings can be very long, trillions of letters, so it is important to find compact representations.  Here we first survey various forms of one potential compaction methodology, the {\it cover} of a given string \s{x}, initially proposed in a simple form in 1990, but increasingly of interest as more sophisticated variants have been discovered.
  We then consider covering by a {\it seed}; that is, a cover of a {\it superstring} of \s{x}.
We conclude with many proposals for research directions that could make significant contributions to string processing in future.
\end{abstract}


\section{Introduction}\label{sec:intro}
A \itbf{string} (as computer scientists call it) or \itbf{word} (as mathematicians call it) is just a sequence $\s{x} = \s{x}[1..n]$ of $n \ge 0$ entries drawn from a finite set of \itbf{letters} called an \itbf{alphabet}, usually denoted by $\Sigma$.
Stringology --- or combinatorics on words --- has existed as a field of scientific enquiry for more than a century, dating back to Axel Thue's foundational paper in 1906 \cite{thue:06}.
Thue showed that infinitely long strings could be constructed, even on a small alphabet, so as to avoid certain simple patterns.
This suggests that it could be difficult, at least in some cases, to extract meaning (represented for example by recurring patterns) from a given text.  Indeed, from its inception, stringology has been deeply concerned with the presence or absence of \itbf{patterns} in strings \cite{S03, CHL14}.

In particular, the recognition \cite{WC53} that the DNA of all organisms is to a first approximation a string on an alphabet $\Sigma = \{a,c,g,t\}$ has over the last 70 years led to a huge and exponentially increasing bioinformatics literature that deals essentially with the application of stringology to fundamental biological research --- the search for biologically significant patterns.

In this survey we follow the study of one of these patterns --- the ``cover'' \s{u} of a string \s{x} --- from its initial definition \cite{AE90} as a ``quasiperiodicity'' (every position in \s{x} lies within an occurrence of \s{u}) to more flexible concepts, 30 years later, defined in terms of a repeating substring \s{u} whose occurrences maximize, over all repeating substrings of \s{x}, the positions covered --- and which, moreover, can be efficiently computed.
More generally, we also discuss methods to find a ``seed'' of \s{x} --- that is, a repeating substring \s{u} of \s{x} that covers a string \s{w} which contains \s{x} as a substring \cite{imp:93,IMP96} --- a more general computation that nevertheless, can often be competitive in terms of execution time.
Thus, potentially, the covers/seeds of a string \s{x} could provide a comprehensive, compact and usefully organized representation of \s{x} that would assist in its interpretation and processing.

For those not familiar with the terminology of strings, Section~\ref{sect:prelim} provides basic definitions.  Then Section~\ref{sect:cover} gives an overview of the advances (and occasional retreats) in the various extensions of the idea of a ``cover'' and their application to some kind of canonical representation of a string.   Section~\ref{sect:seeds} goes on to deal with the often non-trivial generalization of covering algorithms to seeds.  Finally, Section~\ref{sect-open} outlines future research directions that may lead to more useful and precise characterizations of the representative ``patterns'' contained in a given text.

\section{Preliminaries}\label{sect:prelim}

We consider strings $\s{x} = \s{x}[1..n]$, where each entry $\s{x}[i]$ consists of one or more distinct symbols (\itbf{letters}) chosen from a set $\Sigma = \{\lambda_1, \lambda_2, ..., \lambda_{\sigma}\}$ of finite size $\sigma = |\Sigma|$, called an \itbf{alphabet}.  It is often convenient to suppose that the elements of $\Sigma$ are ordered.
Every entry $\s{x}[i]$ that consists of a single letter from $\Sigma$ is said to be \itbf{regular}; if every position in \s{x} is regular, then so is \s{x}.  However, if any $\s{x}[i]$ consists of a set of two or more letters from $\Sigma$, such as $\{a,c,t\}$ or $\{1,3\}$, then both $\s{x}[i]$ and \s{x} are said to be \itbf{indeterminate}.
A \itbf{partial word} is a commonly-occurring special case of an indeterminate string whose entries are either single letters or $\Sigma$ itself (usually termed a \df{hole} and denoted as $*$).
Strings whose indeterminate entries also specify a likelihood that each symbol will occur are called \itbf{weighted} --- these typically occur in bioinformatics applications.  For example, a single letter $\{a,30;c,40;t,30\}$ would indicate that $a/c/t$ is expected to occur $30/40/30$ \% of the time, respectively.

Thus any English-language text is a regular string on some $\Sigma$ consisting of 52 upper and lower case letters, 10 numeric digits, space, and a variety of special symbols --- and so, for English, $\sigma$ may be as much as 90.
On the other hand, an indeterminate string could be a DNA fragment such as $\s{x} = \{a,c\}g\{g,t\}c$ on alphabet
$\Sigma = \{a,c,g,t\}$, some of whose entries ($\{a,c\}, \{g,t\}$) are not well defined or somehow optional.
The results presented in this survey, unless otherwise indicated, are restricted to regular strings.

\medskip
The \itbf{length} of $\str{x}[1..n]$ is $|\str{x}|=n$.
Two regular strings \s{x} and \s{y} of length $n$ are said to \itbf{match} if $\s{x}[i] = \s{y}[i], 1 \le i \le n$\footnote{When at least one of \s{x}, \s{y} is indeterminate, we require for a match only that the intersection $\s{x}[i] \cap \s{y}[i]$ be nonempty.}.
The \itbf{empty string} $\varepsilon$ is a string of length $n = 0$.
A string $\str{x}[i..j]$ is a \df{substring} (or \df{factor}) of $\str{x}[1..n]$ if
$1 \leq i \leq j \leq n$, a \df{proper substring} if moreover $j-i+1 < n$; when \s{x} is regular, $\str{x}[i..j]$ is said to be a \df{repeating substring} if it matches another substring of \s{x} --- that is, if there exists $i' \ne i$ and
$j' \le n$ such that $\s{x}[i'..j'] = \s{x}[i..j]$.
A repeating substring \s{u} is said to be \df{extendible} if all occurrences of \s{u} in \s{x} are followed or preceded by the same letter; otherwise, \df{nonextendible}.
If \s{x} is a proper substring of a string \s{w}, then we say that \s{w} is a \itbf{superstring} of \s{x}.
A \df{prefix} (\df{suffix}) of nonempty $\str{x}$ is a substring $\str{x}[i..j]$, where $i=1$ ($j=n$) --- it is moreover convenient to suppose that every nonempty \s{x} has the empty string as both prefix and suffix.  The suffix starting at position $i$ is sometimes called \df{suffix $i$}.
If $\s{x} = \s{uv}$ for nonempty \s{u}, \s{v}, then $\s{x'} = \s{vu}$ is said to be a \itbf{rotation} of \s{x}.

Given $\s{x} = \s{x}[1..n]$ on alphabet $\Sigma$, the \itbf{Parikh vector} $P = P_{\s{x}}[1..\sigma]$ of \s{x} is an integer array such that, for
$1 \le \ell \le \sigma$, $P[\ell]$ is the number of occurrences of letter $\lambda_{\ell}$ in \s{x}.
A string \s{x} consisting of $r \ge 1$ concatenated copies of a string \s{u} is denoted by $\s{u}^r$.
Then \s{x} is said to be \itbf{periodic} with \itbf{period} $|\s{u}|$ if it can be represented as $\str{x}=\str{u}^r\str{u'}$, where $\str{u'}$ is a (possibly empty) prefix of $\str{u}$, and either $r \geq 2$ and  or else $r = 1$ and $\str{u'}$ is not empty.
In the former case ($r \ge 2$), $\s{u}^r$ is said to be a \itbf{repetition}; any string that is not a repetition is said to be \itbf{primitive}.
A \itbf{border} of \s{x} is a proper prefix \s{u'} of \s{x} that matches a suffix of \s{x}; thus every nonempty string has an empty border.
Note that if \s{x} has a nonempty border \s{u'}, then \s{x} is necessarily periodic with period $|\s{x}|-|\s{u'}|$.
Furthermore, if \s{u'} is a border of \s{x}, then every border of \s{u'} is also a border of \s{x}.
For example, the string $\s{x} = ababaaba$ has borders $aba$, $a$ and $\varepsilon$, hence periods $8-3 = 5$ and $8-1 = 7$; and note that the shorter border $a$ must also be a border of the longer border $aba$.

\medskip
The \itbf{frequency} $f_{\str{x},\str{u}}$ of a substring $\str{u}$ in a string $\str{x}$ is the number of occurrences of $\str{u}$ in $\str{x}$.
In the preceding example, choosing \s{u} to be the border $aba$, we see that $f_{ababaaba,aba} = 3$.
In fact, we observe that {\it every} position in \s{x} lies within an occurrence of $\s{u} = aba$.
We formalize this idea with the definition of a \itbf{cover} of \s{x}; that is, a repeating substring \s{u} in \s{x} such that every position in \s{x} lies within an occurrence of \s{u}.
In such a case we say that \s{x} is \itbf{quasiperiodic}.
From the example $\s{x} = ababababa$, we discover that a string may have more than one cover --- in this case, $\s{u_1} = aba$, $\s{u_2} = ababa$, $\s{u_3} = abababa$ --- such that, moreover, every longer cover \s{u_i} is covered by every shorter one \s{u_{i-1}}, \s{u_{i-2}}, \ldots, \s{u_1}.
It is easy to see that every cover of \s{x} must also be a border of \s{x}; thus the set of covers is a subset of the set of borders, and so contains at most $\bigO(n)$ elements.
The maximum is achieved by $\s{x} = a^n$, which has $n-1$ covers $a^i, 1 \le i \le n-1$; more generally, every repetition $\s{x} = \s{u}^k$ has $k-1$ covers $\s{u}^i, 1 \le i < k$.

As noted in the Introduction, it turns out to be useful to generalize the idea of a cover to that of a \itbf{seed}; that is, a proper substring of \s{x} which is a cover of a superstring \s{w} of \s{x} \cite{imp:93,IMP96}.
Then, for example, the substring $\s{u_1} = aba$ is a seed of $\s{x'} = abababab$ since, as we have just seen, it is a cover of $\s{x} = ababababa = \s{x'}a$, a superstring of \s{x'}.
An important difference between covers and seeds is that the number of seeds of \s{x} may be $\Theta(n^2)$: \cite{KKRRW20} gives the example
$\s{x} = a^mba^mba^mba^m$ of length $n = 4m+3$, whose seeds include the $\Theta(m^2)$ distinct substrings $a^iba^j$ determined by the rule that $i$ and $j$ assume all values $0 \le i,j \le m$ such that $i+j \ge m$.

In Section~\ref{sect:cover} we discuss the various (and successively more sophisticated) kinds of cover for strings \s{x} that have been proposed over the last 30 years and outline the methodology for their computation.  In Section~\ref{sect:seeds} we then go on to discuss the (often surprisingly different) methods proposed for computing seeds.
Many of these methods depend heavily on the data structures surveyed below.

\medskip
We first explain an important extension of the idea of periodicity.
A nonempty substring $\s{v} = \s{x}[i..j] = \s{u}^r\s{u'}$ of \s{x} is said to be a \itbf{run} of \itbf{period} $p = |\s{u}|$ in \s{x} if $r \ge 2$, \s{u'} is a possibly empty proper prefix of \s{u}, and there exist {\it no} integers
$i' \le i, j' \ge j$ such that
\begin{itemize}
    \item [(1)] $\s{x}[i..j]$ is a proper substring of $\s{x}[i'..j']$; and
    \item [(2)] $\s{x}[i'..j']$ has period $p$.
\end{itemize}
Thus a run, denoted $(i,j,p)$, is \itbf{maximal}: any extension left or right of \s{v} in \s{x} yields a substring that is {\it not} a run of period $p$.

\medskip
For example, the string $\s{x} = abacababacabacaba$ contains runs $\s{x}[5..9] = ababa = (ab)^2a$ of period $2$,
$\s{x}[7..17] = abacabacaba = (abac)^2aba$ of period 4, and $\s{x}[1..13] = abacababacaba = (abacab)^2a$ of period 6.
These runs represented as triples are $r_1 = (5,9,2)$,
$r_2 = (7,17,4)$ and $r_3 = (1,13,6)$, respectively.
Note however that $\s{x}[7..16] = (abac)^2ab$
and $\s{x}[10..17] = (caba)^2$ are {\it not} runs because they are not maximal.

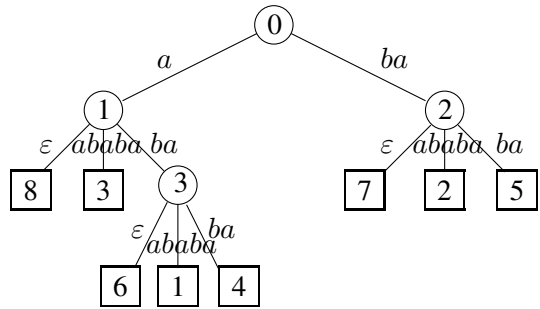
\begin{figure}[!ht]
\vspace*{-2mm}
$$ \begin{array}{rccccccc}
\scriptstyle 1 & \scriptstyle 2 & \scriptstyle 3 & \scriptstyle 4 & \scriptstyle 5 & \scriptstyle 6 & \scriptstyle 7 & \scriptstyle 8 \\
\s{x} = a & b & a & a & b & a & b & a \\
\SA_{\s{x}} = 8 & 3 & 6 & 1 & 4 & 7 & 2 & 5 \\
\LCP_{\s{x}} = 0 & 1 & 1 & 3 & 3 & 0 & 2 & 2 \\
\RSF_{\s{x}} = 0 & 5 & 5 & 3 & 3 & 0 & 3 & 3 \\
\OLP_{\s{x}} = 0 & 0 & 0 & 1 & 0 & 0 & 0 & 0
\end{array} $$
\begin{center}
\setlength{\unitlength}{0.8pt}
\begin{picture}(300,150)(-165,-130)
\put(0,0){\circle{18}}\put(-8.1,-4){\line(-2,-1){63}}\put(8.1,-4){\line(2,-1){63}}
\put(-3,-4){0}
\put(-80,-40){\circle{18}}\put(-86.4,-46.4){\line(-1,-1){22}}\put(-73.6,-46.4){\line(1,-1){22}
}\put(-80,-49){\line(0,-1){19.5}}
\put(-83,-44){1}
\put(80,-40){\circle{18}}\put(73.6,-46.9){\line(-1,-1){22}}\put(86.4,-46.9){\line(1,-1){22}}
\put(80,-49){\line(0,-1){19.5}}
\put(77,-44){2}
\put(-55,-20){$a$}\put(50,-20){$ba$}
\put(33.6,-86.4){\framebox(18,18){7}}
\put(50,-60){$\varepsilon$}
\put(71,-86.4){\framebox(18,18){2}}
\put(65,-60){$ababa$}
\put(105,-86.4){\framebox(18,18){5}}
\put(104,-60){$ba$}
\put(-123,-86.4){\framebox(18,18){8}}
\put(-110,-60){$\varepsilon$}
\put(-89,-86.4){\framebox(18,18){3}}
\put(-95,-60){$ababa$}
\put(-45,-75){\circle{18}}\put(-50,-83){\line(-1,-2){15}}\put(-41,-84.3){\line(1,-2){15}}\put(-45,-84){\line(0,-1){30}}
\put(-47,-78){3}
\put(-58,-60){$ba$}
\put(-81,-131.4){\framebox(18,18){6}}
\put(-67,-100){$\varepsilon$}
\put(-54,-131.4){\framebox(18,18){1}}
\put(-60,-107){$ababa$}
\put(-25.4,-131.4){\framebox(18,18){4}}
\put(-31,-100){$ba$}
\end{picture}
\end{center}\vspace*{-3mm}
\caption{Suffix array, $\LCP/\RSF/\OLP$ arrays and corresponding suffix tree for $\s{x} = abaababa$ --- adapted from \cite{S13}.}
\label{fig-st}\vspace*{-4mm}
\end{figure}

\medskip
A \df{suffix array} $\SA_{\str{x}}$ of $\str{x}$ \cite{MM90,MM93, Karkkainen:2003, KSB06} is an integer array of length $n$, where $\SA_{\str{x}}[i]$ is
the starting position of the $i$-th lexicographically least (lexleast) suffix in $\str{x}$.  (Thus an ordering of the alphabet is required.)  The \df{longest common prefix array} $\LCP_{\str{x}}$ of $\str{x}$ \cite{MM90,MM93, KLAAP01} is an integer array of length $n$, where $\LCP_{\str{x}}[1] = 0$ and
$\LCP_{\str{x}}[i]$, $1< i \leq n$, is the length of the longest common prefix of suffixes starting at positions $\SA_{\str{x}}[i-1]$ and $\SA_{\str{x}}[i]$.
Over the last quarter-century these data structures, both of them now recognized \cite{KLAAP01, Karkkainen:2003, KSB06,S13} to be efficiently computable in $\bigO(n)$ time, have come to be used extensively in a wide variety of string algorithms: taken together they permit equal substrings of \s{x} to be identified (since those substrings will naturally occur close together in the sorted suffix array).
For example, in Figure~\ref{fig-st} the three occurrences of substring $aba$ in $\s{x} = abaababa$ --- starting at positions 1, 4 and 6 --- occur conveniently close together in
$\SA = \SA_{\s{x}}$, identified by $\SA[3..6] = 614$.  Furthermore, since $\LCP=\LCP_{\str{x}}$ is keyed to $\SA$, the value
$\LCP[4] = 3$ tells us that there are equal substrings of length 3 beginning at $\s{x}[\SA[3]] = \s{x}[6]$ and
$\s{x}[\SA[4]] = \s{x}[1]$.
Thus these two integer arrays of length $n$ are powerful computational tools for the analysis of strings --- for, in a sense, determining their ``meaning''.

The $\SA$/$\LCP$ arrays provide compact storage for a more general structure, called the \itbf{suffix tree} of \s{x}, denoted by $\ST = \ST_{\s{x}}$.
As shown in Figure~\ref{fig-st}, $\ST$ is a search tree with $n$ leaf nodes representing the $n$ nonempty suffixes of \s{x}; each edge descending from each internal node represents a distinct letter or substring, which are available in ascending lexicographic order.  For example, the edges descending from node 0 in Figure~\ref{fig-st} correspond to $a$ and $ba$, the only possible prefixes of any suffix of \s{x}.  At node 1, the only possible continuations from $a$ are $\varepsilon$, corresponding to position 8 in \s{x}, $ababa$, corresponding to position 3, and $ba$, corresponding to one of positions $6,1,4$.
Thus a preorder traversal of $\ST_{\s{x}}$ yields the suffixes of \s{x} in ascending lexorder (exactly the entries referenced by $\SA_{\s{x}}$), while the path from the root of the tree to each node spells out the common prefix to all of that node's descendants in $\ST_{\s{x}}$ (exactly the entries in $\LCP_{\s{x}}$).

Like $\SA_{\s{x}}$ and $\LCP_{\s{x}}$,
$\ST_{\s{x}}$ can be computed in $\bigO(n)$ time \cite[Ch.\ 5.2]{S03}, but nevertheless, as a practical matter, its construction is much slower.
  Moreover, the tree structure, due to the requirement for pointers and other auxiliary information, normally consumes much more space in computer memory.
However, suffix trees are still used in many applications because the basic tree can be ``annotated'' in various ways with information useful for processing.  See however \cite{AO04}.

The suffix tree was first proposed and computed almost half a century ago \cite{Weiner73}, the suffix array 30 years ago \cite{MM90}.  Surveys of their computation and use are available in \cite{S13, Apostolico85,PST07}.

For definitions of the $\RSF_{\s{x}}$ (Repeating Substring Frequency) and $\OLP_{\s{x}}$ (Overlapping Positions) arrays, see Subsection~\ref{ssec:oc}.

\section{Covers in strings}\label{sect:cover}
Before discussing the various forms of cover of a string, we pause to say a little more about the border and the period.
Given $\s{x}[1..n]$, the
array $\s{\beta_x}[1..n]$ is said to be the \itbf{border array} of \s{x} if $\s{\beta_x}[i]$ is the length of the longest border of $\s{x}[1..i]$,
$1 \le i \le n$.
Thus for the example $\s{x} = ababaaba$, $\s{\beta_{x}} = 00123123$.  We see that in this case \s{\beta_{x}} provides quite a bit of information about \s{x} --- in fact, with a bit of logic, we can determine the exact structure of \s{x} from $00123123$, which we could characterize using two arbitrary letters $\lambda_1$ and $\lambda_2$: $\lambda_1\lambda_2\lambda_1\lambda_2\lambda_1\lambda_1\lambda_2\lambda_1$. Similarly, the \df{period array} $P_{\s{x}}$ of \s{x} \cite{Christou2013} is an integer array where $P_{\s{x}}[i]$ is the length of the shortest period of $\s{x}[1..i]$, and its dual the \df{suffix period array} $P_{\s{x},suf}$ is defined such that $P_{\s{x},suf}[i]$ is the length of the shortest period of  $\s{x}[i..n]$.
For the above example $\s{x} = ababaaba$,
$P_{\s{x}} = 12222555$, $P_{\s{x},suf} = 55333221$.

Furthermore, \s{\beta} can be elegantly computed in $\Theta(n)$ time using a famous early (1974) string algorithm \cite{ahu-book:74}, giving rise to the tantalizing idea that strings might somehow be organized into useful equivalence classes according to some structural analysis.
Of course, a little reflection tells us that, alas, the border array will not satisfy this requirement: $000000000$ is the border array of $cabaababa$, $caaaaaaaa$, $abcdefghi$, and many other very diverse strings!
In fact, as shown in \cite{AIRSS16}, the expected maximum length of the border of a string on a binary alphabet is only 1.64 letters --- and this value of course decreases precipitously as alphabet size increases.

Closely related to the border array is the \itbf{prefix array} of \s{x}; that is, an array
$\pi = \pi[1..n]$, where $\pi_{\s{x}}[i]$ is the length of the longest substring at position $i$ of \s{x} that matches a prefix of \s{x}.  Thus, in the above example $\s{x} = ababaaba$,
$\pi_{\s{x}} = 80301301$.  It was shown in \cite{BKS13} that, on regular strings, the border array and the prefix array are equivalent in the sense that each can be computed from the other.  However, on indeterminate strings, the prefix array retains its properties --- in particular, identifying all the borders of every prefix --- whereas the border array does not \cite{SW08}.

A recent paper \cite{IR16} by Iliopoulos \& Radoszewski discusses subquadratic solutions to compute both the quantum and deterministic border arrays (see Section~\ref{ssec:coverindet}) and the prefix array of partial and indeterminate strings. They describe $\bigO(n\sqrt{n \log n})$-time algorithms to compute these arrays on partial words. Then they propose $\bigO(n\sqrt{n})$-time algorithms to compute the prefix array and the quantum border array of an indeterminate string over a constant-sized alphabet. They also go on to show that, provided the Strong Exponential Time Hypothesis (SETH) holds, no efficient algorithms exist to compute both the quantum and deterministic border arrays, and prefix array of an arbitrary indeterminate string over a general alphabet.  As we discover below (Section~\ref{ssec:coverindet}), the cover array computation is even more restricted.
These results have recently been put to use in \cite{MS20, DMS22}, where the prefix array, rather than the border array, has been introduced into the Knuth-Morris-Pratt (KMP, \cite{KMP77}) and Boyer-Moore (BM, \cite{BM77}) pattern-matching algorithms, in order to do efficient pattern-matching on indeterminate strings: for text of length $n$, pattern of length $m$, both indeterminate, KMP and BM execute in $\bigO(m\sqrt{m}n)$ time.

\subsection{Covers \& cover arrays}\label{ssec:coverca}

In 1990 Apostolico and Ehrenfeucht \cite{AE90,AE93} introduced the idea of a cover of a string, defined in Section~\ref{sect:prelim}, as a means of providing a description of some strings, at once more succinct and more expressive: $aba$ as a kind of abbreviation of $ababaaba$.
They described an $\bigO(n\log^2n)$ algorithm to compute all the maximal quasiperiodic substrings of \s{x}, in particular \s{x} itself if quasiperiodic, a result later improved by Iliopoulos and Mouchard
\cite{IM99}, also Brodal and Pedersen~\cite{BP00}, to $\bigO(n \log n)$.
In 1991 Apostolico, Farach and Iliopoulos \cite{AFI91} described a linear-time algorithm to determine the shortest cover of quasiperiodic \s{x}. Then Breslauer \cite{B1992} published an on-line linear-time algorithm to compute the minimum cover of each prefix of \s{x}, while Moore and Smyth \cite{MS94,MS95} described a linear-time algorithm to compute {\it all} the covers of \s{x} itself.
A recent paper \cite{CR21} compares the run-time of these latter three algorithms (AFI~\cite{AFI91}, B~\cite{B1992}, MS~\cite{MS94,MS95}), along with implementations of several other generalizations of cover algorithms. Further, in~\cite{IP96} Iliopoulos and Park presented an $\bigO(\log\log n)$-time parallel algorithm to compute all the covers of \s{x}.

\medskip
Finally, Li and Smyth \cite{LS02} published an on-line linear-time algorithm to compute the \itbf{cover array} $\s{\gamma_x}[1..n]$ of \s{x}, specifying the longest cover of each prefix $\s{x}[1..i]$ of \s{x}, zero for no cover.  Since for every cover \s{u} of \s{x}, any cover of \s{u} is also a cover of \s{x}, the  cover array $\s{\gamma}$ turns out to be an exact analogy to the border array $\s{\beta}$, specifying {\it all} the covers of every prefix of \s{x}:
$$ \begin{array}{rccccccccccccc}
\scriptstyle 1 & \scriptstyle 2 & \scriptstyle 3 & \scriptstyle 4 & \scriptstyle 5 & \scriptstyle 6 & \scriptstyle 7 & \scriptstyle 8 & \scriptstyle 9 & \scriptstyle 10 & \scriptstyle 11 & \scriptstyle 12 & \scriptstyle 13 \\
\s{x} = a & b & a & a & b & a & b & a & a & b & a & a & b \\
\s{\beta} = 0 & 0 & 1 & 1 & 2 & 3 & 2 & 3 & 4 & 5 & 6 & \red{1} & 5 \\
\s{\gamma} = 0 & 0 & 0 & 0 & 0 & 3 & 0 & 3 & 0 & 5 & 6 & 0 & 5 \\
\end{array} $$
Thus the cover array, even more precisely than the border array, can describe the structure of certain strings \s{x}.

\medskip
This observation suggests the possibility of
inferring a string that corresponds to a given cover array,
first stated and solved in \cite{CIPT10}:
given a valid cover array $\gamma$ of length $n$, find in linear time a corresponding string \s{x} on an alphabet of minimum size whose cover array is $\gamma$.
Then three years later, in \cite{MNRR13, MNRR13a}, a remarkable linear-time algorithm was described that, for every valid $\gamma$, computes a corresponding string on a two-letter alphabet.

In this context, a recent paper \cite{GKRRW20} shows that \s{x}, though not necessarily on a minimum alphabet, can be determined in linear time from different kinds of repetitions or symmetries present in the string --- in particular, corresponding to a given valid border array, prefix array, or other features.

Unfortunately, far fewer strings have covers than have nonempty borders~\cite{AIRSS16}, so that the range of application of the cover array is correspondingly limited.  Even short strings on a small alphabet rarely have a cover --- for $n = 4$ and $\sigma = 2$, only four of 16 distinct nonempty strings have a cover --- so that for longer strings on a larger alphabet, almost always $\s{\gamma_x}[i] = 0$.

\medskip
Nevertheless, almost 20 years after the publication of the cover array algorithm, two recent papers have refocussed attention on cover computation by describing algorithms to compute covers, first of every rotation, then of every substring, of a given string \s{x}:
\begin{itemize}
\item
In \cite{CCRR20-1, CCRR21-1} Crochemore {\it et al.} first describe an $\bigO(n\log n)$-time algorithm to compute the shortest cover of every rotation of \s{x}, then an $\bigO(n)$-time algorithm to compute the shortest among these covers. This result is further improved by the same authors in \cite{CIR22} where they propose an $\bigO(n)$-time algorithm for the same problem.
\item
In \cite{CCRR20-2} the same authors preprocess \s{x} in $\bigO(n\log n)$ time and space in order to be able to \mbox{compute}, for any selected substring $\s{u} = \s{x}[i..j]$ of \s{x}, both the shortest cover (in time \linebreak
\mbox{$\bigO(\log n \log\log n)$}) and in addition {\it all} the covers (in time $\bigO(\log n (\log\log n)^2)$).
\end{itemize}
These algorithms are complex, requiring not only  computation of $ST_{\s{x}}$, but also of all the seeds of \s{x}.  For this latter calculation, they make use of a linear-time algorithm \cite{KKRRW20}, discussed in Section~\ref{sect:seeds}, that computes a linear encoding (a \itbf{package representation}) of the seeds --- even though, as we have seen, their occurrences may be quadratic in number.

Another recent development has also inspired renewed interest in forms of quasiperiodicity: the generalization of pattern matching and periodicity under a \itbf{Substring Consistent Equivalence Relation} (SCER) denoted by $\approx$.  For strings \s{x} and \s{y}, \cite{MAIBT16} defines
$\s{x} \approx \s{y}$ iff $|\s{x}| = |\s{y}|$ and
$\s{x}[i..j] \approx \s{y}[i..j]$ for every $1 \le i \le j \le |\s{x}|$.  The authors then describe efficient pattern-matching using $\approx$ as well as an analogue of Fine \& Wilf's periodicity lemma \cite{FW65}.  This work is followed up in \cite{KHYS20}, where the cover array is defined for SCERs and existing shortest/longest cover array algorithms are appropriately generalized.

On the other hand, another recent theoretical result provides a basis for understanding that quasiperiodic strings must be in some sense rare: Amir {\it et al.} \cite{AIR17} show that any two distinct quasiperiodic strings of the same length must differ at more than one position (Hamming distance greater than one).

No doubt due to the scarcity of exact covers in strings, research in this area has for the last quarter century focused on various forms of multiple or approximate cover, as discussed below.

\subsection{$k$/$\lambda$-covers}\label{ssec:kcover}

In 1998 Iliopoulos and Smyth \cite{IS98} introduced the \itbf{$k$-cover} problem: for given \s{x} and $k > 1$, determine a covering set $U_{t,k} = \{\s{u_1},\s{u_2},\ldots,\s{u_t}\}$ of $t$ substrings of \s{x}, each of length $k$, such that every position of \s{x} lies within some element of $U_{t,k}$.
For given $k$, let $\tau_k$ denote the minimum value of $t$ (if it exists) for which $U_{t,k}$ is a covering set; in this case $U_{\tau_k,k}$ is said to be a \itbf{minimum $k$-cover} of \s{x}.
For example, given $\s{x} = abaabbaaab$ and $k = 2$, $U_{3,2} = \{ab,aa,bb\}$ and $\{ab,ba,aa\}$ are minimum 2-covers of \s{x} with $\tau_2 = 3$, while
$U_{3,3} = \{aba,aab,baa\}$ and $\{aba,aab,bba\}$
are minimum 3-covers with $\tau_3 = 3$.
Observe that every $k$-cover must include the suffix and the prefix of length $k$; thus, whenever \s{x} has no border of length $k$, $\tau_k \ge 2$ (if it exists).
Of course of particular interest is the smallest value of $k$ that yields a minimum $k$-cover corresponding to the smallest $\tau_k$.
Thus in the above example, we would prefer $U_{3,2}$ to $U_{3,3}$.

Also in \cite{IS98} it was ``proved'' that, for given \s{x}, the minimum $\tau_k$ could be computed in polynomial time, a result later found to be incorrect: in 2005 Cole {\it et al.} \cite{CIMSY05} showed that the problem of determining the cardinality $\tau_k$ of a minimum $k$-cover is NP-complete for every $k \ge 2$.
Nevertheless, in the same paper \cite{CIMSY05}, the authors described two $\bigO(n\log n)$ algorithms that approximated the minimum $k$-cover $U_{\tau_k,k}$ to within a logarithmic ($\log n$) factor.  Then in 2011 (Iliopoulos {\it et al.} \cite{IMS11}) presented a polynomial-time algorithm to approximate $U_{\tau_k,k}$ to within a factor $k$.
For more on this topic see below, Section \ref{ssec:approxc}.

In \cite{GZI06a, GZI07}, summarized in \cite{ZGI08}, Guo {\it et al.} introduced the \df{$\lambda$-covers} problem, a parameterized version of $k$-covers: for given positive integers $k$ and $\lambda$, find if possible a set $S = S_{k,\lambda}$ containing exactly $\lambda$ substrings of \s{x}, each of length $k$, such that the entries in $S$ cover \s{x}.
For an alphabet of size $\sigma$, the authors presented an $\bigO(\sigma k n^2)$ algorithm that solved the problem for values $1,2,\ldots,k$ and fixed $\lambda$.
To facilitate an efficient solution,
rather than suffix trees,
the authors made use of the Equivalence Class and Reversed Equivalence Class Trees (ECT, RECT), introduced in \cite{IPZ04}, to compute $S_{k,\lambda}$ whenever possible, still in $\bigO(n^2)$ time. However, unfortunately, these results turn out to be incorrect, to an extent that has not yet been made fully precise: in \cite[p.\ 26]{CR21} Czajka \& Radoszewski describe a counterexample which makes clear that the $\lambda$-covers, as defined in these papers, cannot be computed in $\bigO(n^2)$ time.

In \cite{RS20} Radoszewski \& Straszy\'{n}ski consider the special case $\lambda = 2$ of parameterized $k$-covers: they describe an algorithm requiring $\bigO(n \log^{1+o(1)}n)$ time that, over all $k$, specifies all the {\it pairs} of strings of length $k$ that cover $\s{x}[1..n]$.  Their algorithm generalizes to $\lambda > 2$, but with an order $n$ multiplicative increase in complexity for each increase in $\lambda$.

\subsection{$\alpha$-partial cover}\label{ssec:alphac}

In \cite{KRRPW2015} Kociumaka
{\it et al.} introduce the  \itbf{cover index}
$C = C(\s{x},\s{u})$ of a string \s{u} within \s{x}; that is, the number of positions in \s{x} covered by a repeating substring \s{u}, which is therefore called a \itbf{partial cover} of \s{x}.
For example, given $\s{x} = abaababaabaab$ of length $n = 13$, $C(\s{x},aba) = 11$ and $C(\s{x},ab) = 10$
are the cover indices for partial covers $aba$ and $ab$, respectively.
For a given integer $\alpha \in 1..n$, the authors describe an algorithm, executing in $\bigO(n\log n)$ time and $\Theta(n)$ space, to compute all shortest substrings \s{u} of \s{x} such that $C(\s{x},\s{u}) \ge \alpha$.
Thus for the same example \s{x} and $\alpha = 10$, the \itbf{$\alpha$-partial-cover} algorithm would return $ab$, the shortest substring covering at least 10 positions.
For $\alpha = 11$, it would return $aba$.
Only for the choice $\alpha = 15$ would it return the full cover $abaab$ of \s{x}.

However, to correct this deficiency, \cite{KRRPW2015} goes on to describe the \itbf{all-partial-covers} algorithm, with the same asymptotic time and space requirements, which computes the $\alpha$-partial cover for {\it all} $\alpha \in 1..n$.

This approach was the first to enable computation of a cover of \s{x} that is in some sense optimal without requiring that it also be a border.  The price paid for this achievement is that the algorithm requires the use of the \textit{cover suffix tree} -- a suffix tree augmented with additional nodes and additional values at each node -- thus increasing both time and space requirements. This algorithm, among others, was implemented in~\cite{CR21}. More recently, Radoszewski in~\cite{radoszewski2023} proposes a remarkable linear time construction of the cover suffix tree data structure - thus showing that the all-partial-covers of a given string can be computed in $\bigO(n)$-time.


In \cite{FIKPPST13} Flouri {\it et al.} introduced a restricted version of the partial cover called the \itbf{enhanced cover}: that is, a border (rather than a repeating substring) of \s{x} that, over all  borders,
covers a maximum number of positions in \s{x}.
The authors showed that the enhanced cover could be computed in $\Theta(n)$ time on regular strings.
Alatabbi {\it et al.} \cite{AIRSS16} extended this idea to the \itbf{minimum enhanced cover} (MEC); that is, the shortest border yielding maximum coverage.
For example, the borders $aba$ and $ababa$ are both enhanced covers of $\s{x} = ababaaabababa$, covering all but one position, but $aba$ is the MEC.
In \cite{AIRSS16} an algorithm was proposed that, according to tests, performed somewhat faster in practice, and that moreover, by using the prefix array rather than the border array for computation, extended the cover calculation to indeterminate strings --- resulting in an average-case $\bigO(n\log n)$, worst-case $\bigO(n^2)$ algorithm.
See also Section~\ref{ssec:lrseed} for a discussion of the \itbf{enhanced seed}.

\subsection{Maximal cover}\label{ssec:oc}
Let $M = M_{\s{x}}$ denote the maximum number of positions in \s{x} covered by {\it any} repeating substring of \s{x}.
Similar to the $\alpha$-partial cover described in Section~\ref{ssec:alphac},
\cite{MSb17} computes a
\itbf{maximal cover}\footnote{There called ``optimal'' cover.} \s{u_M}, a repeating substring of \s{x} that covers $M_{\s{x}}$ positions.
Since \s{u_M} may not be unique, the algorithm can return the longest or shortest \s{u_M}, as required.
Thus, if $C(\s{x},\s{u})$ denotes the number of positions covered by \s{u}, then
$M_{\s{x}} =  C(\s{x},\s{u_M})$.
Note that $M$ and \s{u_M} can also be identified by the $\bigO(n\log n)$ all-partial-covers algorithm of Kociumaka {\it et al.} \cite{KRRPW2015}, described in the previous subsection.

The methodology of \cite{MSb17}  avoids the use of suffix trees, employing instead only simple sorting and runs.  Central to this approach are the $\RSF$ and $\OLP$ arrays, introduced in \cite{MSa17} and \cite{MSb17}, respectively.

\medskip
The $\RSF_{\s{x}}$ (\itbf{Repeating Substring Frequency}) array is defined as follows: for $1 \le i \le n$, $\RSF[i]$ is the frequency in $\s{x}[1..n]$ of the repeating substring of length $\LCP[i]$ that occurs (at least) at the two positions $\SA[i-1]$ and $\SA[i]$; that is, the repeating substring
$$\s{x}[\SA[i], ..., \SA[i] + \LCP[i] - 1].$$
Thus, in Figure~\ref{fig-st},
$\RSF[4] = \RSF[5] = 3$, together with $\RSF[[3] < 3$ and $\RSF[6] < 3$, tells us that the substring, say \s{u}, of length $\LCP[4] = \LCP[5] = 3$  occurs exactly $\RSF[4] = 3$ times altogether in \s{x} at positions
$\SA[3] = 3, \SA[4] = 6, \SA[5] = 1$.
In \cite{MSa17} a simple algorithm is described that computes $\RSF_{\s{x}}$ in $\Theta(n)$ time and space.

\medskip
Consider a repeating substring $\s{u} = \s{x}[\SA[i]..\SA[i] + \LCP[i] - 1]$ in \s{x} of length $\ell = \LCP[i] > 1$ that occurs $k > 1$ times as identified by the maximal sequence
$\LCP[i-1], \LCP[i], ..., \LCP[i+k-1]$.
Then the \itbf{Overlapping Positions} array $\OLP = \OLP_{\s{x}}[1..n]$ specifies, in entry $\OLP[i]$, the total number of overlapping positions (overlaps) between \itbf{consecutive} occurrences of \s{u} in \s{x} --- where positions $\SA[i-1], \SA[i], ..., \SA[i+k-1]$ are not in general in ascending sequence and so must somehow be ordered.
For example, in Figure~\ref{fig-st},
$\OLP[4] = 1$ tells us that the $\RSF[4] = 3$ occurrences of the substring \s{u} at positions $\SA[3..5] = 6,1,4$ of length $\ell = \LCP[4] = \LCP[5] = 3$ --- that is, $aba$ --- have exactly one position ($i = 6$) of overlap in \s{x}.

In \cite{MSb17} two algorithms are described to compute $\OLP_{\s{x}}$, one requiring $\bigO(n\log n)$ time, which has proved to be incorrect \cite{KMS23}, and another requiring $\bigO(n^2)$ time.
In \cite{KMS23} two additional  $\OLP$ algorithms are described, both requiring $\bigO(n^2)$ time in the worst case, but both shown, based on experimental evidence, to execute in linear time in the average case.
Apart from $\OLP$, all other  computations needed to compute the maximal cover require worst-case linear time.

In \cite{GKMS21} Golding
{\it et al.} apply an early $\bigO(n^2)$ implementation MAXCOVER of the maximal cover algorithm to protein sequences; surprisingly, they find significant compression in certain cases.
Further, a version of MAXCOVER, slightly modified to compute nonextendible repeating substrings, is compared to existing software for this purpose, again in the context of protein sequences, and shown to be an order-of-magnitude faster.

\subsection{Frequency cover}\label{ssec:fc}
In \cite{MSa17} a similar, slightly less general, approach to the definition of cover was taken, yielding an algorithm with linear requirements for usage of both time and space.

First identify in \s{x} the set $U = U_{\s{x}}$ of repeating substrings that are not single letters, that occur a maximum number $M$ of times in \s{x}, and that cannot be extended to left or right.
Then the \itbf{frequency cover} \s{u} of \s{x} is defined to be the longest of the entries in $U$.  Thus, for $\s{x} = abaababaabaab$, $U = \{ab\}$ because $ab$ occurs $M = 5$ times in \s{x}, more than any other substring, and so $\s{u} = ab$ --- the same substring chosen by the $\alpha$-Partial Cover algorithm (Section~\ref{ssec:alphac}) for $\alpha = 10$.

From this example, we see that the frequency cover may not in fact be the substring that covers a maximum number of positions ($aba$ of length 3 is not in $U$ but covers 11 positions).
Furthermore, not all strings have a frequency cover (for example, $\s{x} = abc$ or $abaca$), while some strings have multiple frequency covers, perhaps with different properties ($\s{x} = ababacbab$ gives rise to $\s{u_1} = aba$ covering five positions and $\s{u_2} = bab$ covering six positions).

In order to compute the frequency cover efficiently, it turns out to be convenient to make use of the $\RSF$ (Repeating Substring Frequency) array for \s{x}, defined in Section~\ref{ssec:oc}.
For example, in Figure~\ref{fig-st}
$\RSF[2] = 5$ tells us that the substring of length $\LCP[2] = 1$ occurring at position $\SA[2] = 3$ --- that is, $a$ --- occurs 5 times in \s{x}.
Similarly, since $\RSF[5] = 3$, we know that the substring of length $\LCP[5] = 3$ occurring at position $\SA[5] = 4$ --- that is, $aba$ --- occurs 3 times in \s{x}.

\subsection{Approximate covers}\label{ssec:approxc}
These algorithms generally depend on counting the minimum number of \itbf{edit operations} (insertion, deletion or substitution of a single letter) required to transform one string \s{x} into another string \s{x'}, where these operations may have different weights associated with them.
In the following examples, exactly one edit operation transforms $\s{x} \rightarrow \s{x'}$ (and of course {\it vice versa}), implying that the ``distance'' between \s{x} and \s{x'} is one:
\begin{itemize}
    \item{\itbf{insertion}:}
    Insert $c$ at position 2 of $\s{x} = ab$ to form $\s{x'} = acb$;
    \item{\itbf{deletion}:}
    Delete $c$ at position 2 of $\s{x} = acb$ to form $\s{x'} = ab$;
    \item{\itbf{substitution}:}
    Change $c$ to $b$ at position 2 of
    $\s{x} = ac$ to form $\s{x'} = ab$.
\end{itemize}
More generally, given strings \s{x} and \s{x'}, we define \itbf{edit distance ($E$)} $E(\s{x},\s{x'})$ to be the minimum number of edit operations, weighted edit distance ($WE$) if different weights are assigned to different edit operations, \itbf{Levenshtein distance ($L$)}  $L(\s{x},\s{x'})$ the minimum number of insertions and deletions, and \itbf{Hamming distance ($H$)} $H(\s{x},\s{x'})$ the minimum number of substitutions --- required to transform \s{x} into \s{x'} (and {\it vice versa})\footnote{In the literature these definitions vary.  Of course deletion/insertion at position $i$ is just a ``substitution'', so both unweighted edit distance and Hamming distance merely count two distinct Levenshtein operations as one.}.  We use $D$ to indicate any one of $E,L,H$.
For details see \cite[Sect.\ 2.2]{S03}.

\medskip
The idea of an approximate cover of a string was apparently introduced by Sim {\it et al.} \cite{SPKL02} (not available in English), then by Zhang and Blanchet-Sadri \cite{ZB05}, whom we follow here. Given a string \s{x} of length $n$ and a set $U_{t,k} = \{\s{u_1},\s{u_2},\ldots,\s{u_t}\}$ of $t$ strings of identical length $k < n$,  $U_{t,k}$ is said to be a \itbf{$d$-approximate $k$-cover} of \s{x} for some integer $d \ge 0$ if there exists a set $V = \{\s{u_1},\s{u_2},...,\s{u_r}\}$ of $r$ distinct nonempty strings, not necessarily of equal length, such that
\begin{itemize}
\item
$V$ is a cover of \s{x};
\item
for every $\s{u} \in U_{t,k}$, there exists $\s{v} \in V$ such that $D(\s{u},\s{v}) \le d$;
\item
for every $\s{v} \in V$, there exists $\s{u} \in U_{t,k}$ such that $D(\s{u},\s{v}) \le d$.
\end{itemize}

The authors of \cite{ZB05} then described polynomial-time algorithms that, using the results from~\cite{IS98}, compute, for each distance measure $D$\footnote{In \cite{ZB05} the authors define Edit Distance as Levenshtein distance defined here and \textit{vice versa}.}, the minimum integer $d$ such that $U_{t,k}$ is a set of $d$-approximate $k$-covers of \s{x}.  Unfortunately, this result was incorrect:
it was later shown in \cite{CIMSY05} that to determine whether or not any given set $V$ was indeed a minimum $k$-cover was NP-complete (see Section \ref{ssec:kcover}).
Nevertheless, practical algorithms were proposed to compute approximations of $V$   (\cite{ZB05}, \cite{IMS11}), and, as discussed below, several variants of $V$ have been proposed.

Although not a cover problem, in 2001 Sim {\it et al.} \cite{SIPS01} introduced a related and more tractable problem.  Given strings $\s{x}[1..n]$ and $\s{u}[1..m]$, $m < n$, for $D = E,H$ consider partitions
$\s{x} = \s{u_1}\s{u_2} \cdots \s{u_r}$ such that $D(\s{u_i},\s{u}) \le t$, $1 \le i < r$, and $D(\s{u_r},\s{u'}) \le t$ for some prefix \s{u'} of \s{u}.  For each such partition,
$p = |\s{u}|$ is said to be a \itbf{$t$-approximate period} of \s{x}.  A polynomial-time algorithm is described to compute a minimum integer $t$ for which such a partition exists.

In 2005 Christodoulakis {\it et al.} \cite{CIPS05} studied a related cover problem: given strings $\s{x}[1..n]$ and  $\s{u}[1..m]$, $m < n$, consider arrangements of copies of \s{u} placed so as to overlay all positions in \s{x}.  Over all copies of \s{u} in each arrangement, determine the total distance $D$ resulting from mismatches; then an arrangement that minimizes $D$ is a  $\s{u}$-\itbf{approximate cover} of \s{x}.  For $D = H/E$$/WE$\footnote{In \cite{CIPS05} the authors use Edit Distance and Levenshtein distance interchangeably.}, the authors describe algorithms to compute all $\s{u}$-approximate covers in \s{x} requiring time $\bigO(mn)/\bigO(m(n + m))/\bigO(mn^2 + n^2)$, respectively.


More recently, in 2019 Amir {\it et al.} \cite{ALLP19} introduced the \itbf{Approximate Cover Problem (ACP)}: find a ``best'' approximate cover of a given string \s{x} of length $n$; that is, identify a string \s{y} of length $m < n$ whose copies cover a string \s{z}, also of length $n$, in such a way that, over all choices of \s{y}, $H(\s{x},\s{z})$ is minimized.
By considering a relaxed version of ACP, they show that ACP itself is NP-hard. They then discuss two relaxations of the ACP problem, where either a partial or a full ordered list of occurrences of the possible cover of \s{x} is given. They show that both these problems have polynomial time solutions.

In \cite{ALLLP19} the \itbf{Cover Recovery Problem (CRP)} is introduced: given a string $\s{x'} = \s{x'}[1..n]$ that results from the \itbf{approximate} covering of $\s{x}[1..n]$ by an unknown string $\s{u}$ of known length $m$, output a ``small size set'' $S$ of strings of length $m$ such that $\s{u} \in S$.
In \cite{ALP18} further results are presented to assist in understanding the overall complexity of ACP.

Then in 2019, Guth \cite{G19}, building on previous work \cite{GMB08,G16}, introduced a relaxed version of the enhanced cover (Section~\ref{ssec:alphac}). Given a non-negative integer $k$, a \itbf{$k$-approximate enhanced cover} ($k$-AEC) \s{y} of \s{x} is a border of \s{x} such that the total number of positions covered by approximate occurrences of \s{y} in \s{x} exceeds those covered by approximate occurrences of any other border of \s{x}.
Computation of all $k$-AECs is shown to require $\bigO(n^2)$ time.
A ``relaxed'' $k$-AEC is also considered, computable in time $\bigO(n^3)$.

In \cite{KR20, KR21}, given strings
$\s{x} = \s{x}[1..n]$ and \s{u}, an integer
$k \in 0..n-1$, and distance measure
$D$, Kedzierski \& Radoszewski study the \itbf{$(D,k)$-coverage of \s{u} in \s{x}}; that is, the number of positions of \s{x} that lie within a substring \s{v} such that $D(\s{u},\s{v}) \le k$.  For given $k$, they describe an $\bigO(n^2)$-time algorithm to compute $(H,k)$-coverage for all substrings \s{u} of \s{x}, $\bigO(n\log^{1/3}nk^{2/3}\log k)$ for all prefixes.
For $D = E,WE$, they describe algorithms to compute $(D,k)$-coverage for all substrings \s{u}
\eject
\noindent in time $\bigO(n^3)$ and $\bigO(n^3\sqrt{n\log n})$, respectively.
They also show that it is NP-hard to check whether or not a given \s{x} has a $k$-approximate cover (or seed) of given length $\ell$, even on a binary alphabet.

It is noteworthy that,
with the exception of the approach of Guth \cite{G19}, the methodologies described in this subsection avoid the requirement that the approximate cover should be a border of \s{x} --- which as we have seen has an average length of at most 1.64.  No doubt the maximal cover of most strings --- for example, $\s{u} = aba$ for $\s{x} = acabaababaac$ --- will be unrelated to any border.

\subsection{2-Dimensional covers}
\label{ssec:2dc}
In 1996 Iliopoulos \& Korda \cite{CK96} described an
$\bigO(\log \log n)$-time parallel algorithm on the CRCW PRAM model to determine whether a given $n \times n$ square matrix $T$ is superprimitive --- that is, whether there exists a square submatrix $S$ of $T$ such that every position in $T$ lies within an occurrence of $S$; in other words, such that $S$ is a (2D) \itbf{cover} of $T$.  If so, then they return a smallest cover $S$.
Then in 1998 Crochemore {\it et al.} \cite{CIK98} showed how to compute all the covers of a given square matrix $T$ by presenting an $\bigO(n^2)$ time algorithm to compute all square submatrices $P$ of $T$ that cover $T$ --- a result based on the Aho–Corasick~\cite{AC75} automaton and “gap” monitoring techniques.

More recently the problem has been generalized from square to $m \times n$ rectangular matrices $T$, with $N \equiv mn$: in 2019 Popa \& Tanasescu \cite{PT19} describe an average-case $\bigO(N)$-time algorithm to compute a smallest 2D cover by rectangular submatrices $S$ of a given $T$, as well as a worst-case
$\bigO(N^2)$-time algorithm to compute all 2D covers of $T$.  They propose applications such as extraction of textures from images, as well as to image compression and crystallography.

\medskip
A very recent paper \cite{CRRWZ21} by Charalampopoulos {\it et al.} considers two forms of cover of a given $m \times n$ matrix $T$: the 2D cover described above and a 1D cover by a vector $S$ whose occurrences in $T$ are considered both vertically and horizontally.  They present several new results:
\begin{itemize}
\item The smallest 2D cover can be computed in time $\bigO(N)$.
\item All 2D covers can be computed in time $\bigO(N^{4/3})$.
\item All 1D covers can be computed in time $\bigO(N)$.
\end{itemize}

In~\cite{RRSJTZ22} Radoszewski {\it et al.} propose another form of cover called the \df{tile cover}; that is, a string $S$
that covers $T$ by \itbf{non-overlapping} instances of $S$ or its transpose $S^T$. The authors consider two forms of tile cover: 2D-string and 1D-string tile cover (this differs from the 1D cover proposed in \cite{CRRWZ21} by disallowing overlaps). They propose an $\bigO(N)$-time algorithm to compute all 1D tile covers of $T$, and an $\bigO(N^{1+\epsilon})$, $\epsilon > 0$, algorithm to compute all 2D-tile covers of $T$.

\subsection{Specialized covers}
\label{ssec:specialc}

In \cite{ACDGIKW20} Alzamel {\it et al.} introduce the
\itbf{$k$-anticover} of given $\s{x} = \s{x}[1..n]$; that is, for a given integer
$k \ge 2$, a set $S = S_k$ of increasing positions $i$ in \s{x} identifying substrings $\s{x}[i..i+k-1]$, constrained to be distinct, such that every position in \s{x} is contained in an entry from $S$ --- in other words, such that $S$ ``covers'' \s{x}.
For example \cite{ACDGIKW20}, given
$\s{x} = abbbaaaaabab$ of length $n = 12$ and $k = 3$,
$S_3 = \{1,3,5,8,10\}$ is a $3$-anticover of \s{x}, identifying distinct substrings $abb,bba,aaa,aab,bab$ that cover \s{x}; on the other hand, no string $\s{x} = \s{u}\s{v}\s{u}$ has a $|\s{u}|$-anticover.
It is shown in \cite{ACDGIKW20} that for $k \ge 3$ it is NP-hard to determine whether or not a $k$-anticover of \s{x} exists, while a polynomial-time solution exists for $k = 2$.
In \cite{ABK20} Amir {\it et al.} introduce three variants of the $k$-anticover problem, as follows:
\begin{itemize}
    \item \itbf{MaxkAnticover}: find a set $S_k$ that {\it maximizes} the number of covered positions in \s{x};
    \item \itbf{MinRepkAnticover}: if there exists no $S_k$, then find a set $S'_k$ of $k$-strings that allows duplicates and covers \s{x} with a minimum number of repeats of any one entry;
    \item \itbf{MinAnticover}: find the smallest $k$ such that there exists a $k$-anticover of \s{x}.
\end{itemize}
All of these variants are also shown to be NP-hard; however, polynomial-time approximation algorithms are described for each.

\medskip
In \cite{MIBT14} Matsuda {\it et al.} introduce the \itbf{Abelian cover}; that is, a $k$-cover of \s{x} in which each entry has the same Parikh vector  $P = P_{\s{x}}$.
For example, for $k = 3$, $\s{x} = abaab$ has Abelian cover $(aba, aab)$, each with Parikh vector
$P = (2,1)$.  They describe an $\bigO(n)$-time algorithm to compute the longest Abelian cover, whenever it exists, of given $\s{x} = \s{x}[1..n]$, as well as an $\bigO(n^2)$-time algorithm to compute an $\bigO(n^2)$ representation of {\it all (possibly exponential)} Abelian covers of \s{x}.

Similarly, in \cite{GIJLSZ23}, Grossi {\it et al.} introduce the \itbf{cyclic cover} of \s{x}; that is, any substring \s{u} whose rotations cover \s{x}.  In the above example $\s{x} = abaab$, therefore, every rotation of $aba$ is a cyclic cover of \s{x}.  The authors describe an $\bigO(n\log n)$ time algorithm to compute all the cyclic covers of a string.  A recent paper \cite{IKRRWZ23} improves the time requirement for this problem to $\bigO(n)$.

Recall that for $1 \le i \le j \le n$,
$\s{u} = \s{x}[i..j]$ is a \itbf{palindrome} at \itbf{centre} $(i+j)/2$ with \itbf{radius} $(j-i+1)/2$ if $\s{x}[i+h] = \s{x}[j-h]$ for every $h = 0,1,\ldots,(j-i)/2$ --- \itbf{maximal} if there exists no palindrome of greater radius at the same centre.
For given $k \in 1..n$, we say that \s{x} has a \itbf{palindromic cover} $PC_{\s{x},k}$ of \itbf{size} $k$ if every position of \s{x} lies within a palindrome of radius $k/2$.
Of course every single entry $\s{x}[i]$ is a palindrome of radius $1/2$, and so \s{x} always has palindromic cover $PC_{\s{x},1}$.
In \cite{ISIBT14} I {\it et al.} describe a $\bigO(n)$-time and space algorithm to compute the
\textit{smallest} $k$ such that $PC_{\s{x},k}$ is a palindromic cover of given \s{x}.

In~\cite{RRSWZ21b}, Radoszewski {\it et al.}, study covers in both directed and undirected labeled trees. They propose an $\bigO(n\log n/\log \log n)$-time algorithm to compute all covers of a directed (rooted) tree, and an $\bigO(n^2)$-time and space algorithm to compute all covers of an undirected labeled tree.

Recently, in~\cite{CPRRWZ22} Charalampopoulos {\it et al.} introduce the \df{subsequence cover (or s-cover)} of \s{x}; that is, a substring \s{u} whose occurrences as subsequences cover all the positions in \s{x}. They present a linear-time algorithm to test whether a given string \s{v} is an s-cover of a word \s{x}, where \s{x} is defined on polynomially-bounded integer alphabet. They then present a $\bigO(n)$-time algorithm to compute the shortest s-cover of the given \s{x}, where \s{x} is defined on a constant sized alphabet.

\subsection{Extensions to indeterminate \& weighted strings}
\label{ssec:coverindet}
Substantial work has been done on extending covering algorithms to indeterminate strings.

Of course every cover is a border: we have noted above \cite{AIRSS16} that the {\it expected length} of the maximum border of a regular string does not exceed 1.64.  For partial words Iliopoulos {\it et al.} showed \cite{IMMPST02} that the {\it expected number} of borders was less than 3.5; for indeterminate strings Bari {\it et al.} \cite{BRS09} showed that this quantity was less than 29.1746.

In 2003 Iliopoulos {\it et al.} \cite{IMMP03} described two algorithms to compute the border array of a partial word, both requiring $\bigO(n^2)$ time in the worst case, $\bigO(n)$ time on average.
In the same year \cite{HS03} Holub \& Smyth described border array calculations with the same quadratic time complexity on both partial words and indeterminate strings.  Holub \& Smyth also make a distinction between \itbf{quantum} borders, which allow indeterminate letters to match in more than one way, and \itbf{deterministic} borders, in which only a single match is allowed.  They give the example $\s{x} = a*\!*c$, which has two quantum border pairs, $(a*,*c)$ and $(a**,**c)$, requiring $\s{x}[2]$ to match both $c$ and $a$, but only one deterministic border pair --- either $(a*,*c)$, requiring $\s{x}[2] = c$, or the pair  $(a**,**c)$, requiring $\s{x}[2] = a$.


In 2008 Antoniou {\it et al.} \cite{ACIJL08} introduce the idea of a \itbf{$q$-conservative} indeterminate string --- that is, a string \s{x} containing at most $q \ge 0$ indeterminate letters.
They suggest use of the Aho-Corasick automaton \cite{AC75} to determine whether, for given nonnegative integers $q$, $q'$ and $m$, $q$-conservative \s{x} has a $q'$-conservative cover of length  $m$. There is no clear description of an algorithm.  A subsequent paper \cite{AIJR08} discusses covers of DNA strings on $(a,c,g,t)$.

In 2009 Bari {\it et al.} \cite{BRS09} present an {\it average-case} $\bigO(n)$-time algorithm, also using the Aho-Corasick automaton, to compute all the covers of a regular or indeterminate string \s{x} based on the  border array algorithm in \cite{HS03}.  (For indeterminate strings, the cover also may be indeterminate.)  They extend their algorithm to compute the cover array in $\bigO(n^2)$ time.

More recently, Crochemore {\it et al.} in~\cite{CIKRRW17} consider the problem of finding a shortest \itbf{regular} cover of an indeterminate string \s{x}, showing that the computation is NP-complete even over strings \s{x} that are restricted to partial words.  However, they also describe ``near-optimal'' FPT (Fixed Parameter Tractable) algorithms for both partial words and the general case, based on knowledge of a parameter $k$ --- the number of non-regular letters in \s{x}.

\medskip
Weighted strings introduce a new form of ambiguity, as the following example, taken from Zhang {\it et al.} \cite{ZGI10}, demonstrates.  In the weighted string
$$ \s{x} = (a,50;c,25;g,25) g (a,60;c,40) (a,25;c,25;g,25;t,25) c,$$
the pattern $\s{u} = agc$ matches two overlapping substrings:
$$\s{x}[1..3] = (a,50;c,25;g,25)g(a,60;c,40)$$
and $$\s{x}[3..5] = (a,60;c,40)(a,25;c,25;g,25;t,25) c$$
with probabilities $p_1 = 0.5 \times 1 \times 0.4 = 0.2$ and
$p_2 = 0.6 \times 0.25 \times 1 = 0.15$, respectively.
In the first of these,
$\s{u}[3] = c$ matches with $\s{x}[3]$, but in the second,
$\s{u}[1] = a$ provides the match with $\s{x}[3]$.
If overlapping matches that depend on this ambiguous use of an indeterminate letter are allowed, then the matching is called \itbf{loose}; if not, then   \itbf{strict}.
(See above, quantum/deterministic.)
In \cite{ZGI10} the authors outline  $\bigO(n^2)$-time algorithms to compute all the covers of weighted strings, using both  loose and strict matching.

\medskip
An $\bigO(n)$-time algorithm to compute the covers of a weighted string is also given in \cite{IMPPTT06}, based on prior calculation of a \itbf{weighted suffix tree} in time
$\bigO(\sigma n)$.  A recent paper \cite{BKLPR19} describes a more efficient cover calculation based on a ``weighted index''.

\section{Seeds of strings}
\label{sect:seeds}
As defined earlier, the \df{seed} of a string \s{x} is a proper substring \s{u} of \s{x} that is a cover of a superstring $\str{w}$ of $\str{x}$. The notion of seed was first introduced in \cite{imp:93, IMP96} by Iliopoulos {\it et al.}, where they describe an $\bigO(n \log n)$-time algorithm to compute all the seeds of a given string $\str{x}[1..n]$, by computing a linear representation of seeds. Berkman {\it et al.} in~\cite{BIK95} present a parallel algorithm to compute all seeds in $\bigO(\log n)$ time and
$\bigO(n^{1+ \epsilon})$ space,
using $n$ processors in the CRCW PRAM model. Then in~\cite{Christou2013} Christou {\it et al.} present an alternate sequential $\bigO(n\log n)$ algorithm for computing the shortest seed of a given string. In \cite{S00} Smyth poses the question whether all seeds for the given string can be computed in time linear in its length. In 2012 Kociumaka {\it et al.} \cite{KKRRW12} answer this question in the affirmative by  presenting the first linear time algorithm to compute seeds --- though based on the assumption of an integer alphabet. Their algorithm was complex and required constructing a representation of seeds on two suffix trees. In 2020, the same authors \cite{KKRRW20} present a solution to the same problem that uses a much simpler approach called the \itbf{package representation} --- again based on an integer alphabet. The authors define a \itbf{package} to be a collection of consecutive prefixes of a substring of \s{w}; that is, a package is defined as
$$pack(i,j_1,j_2) = \{w[i..j]: j_1 \leq j \leq j_2\},$$
where $i\leq j_1 \leq j_2 \leq |\s{w}|$. If $\mathcal{L}$ is a set of ordered integer triples, they then define
$$PACK(\mathcal{L}) = \bigcup_{(i,j_1,j_2)\in \mathcal{L}} pack(i,j_1,j_2)$$

Their solution outputs the set $\mathcal{L}$ such that the seeds of \s{w} are exactly the elements of  $PACK(\mathcal{L})$.  Furthermore, all packages in the representation are pairwise disjoint; that is, each seed belongs to exactly one package. It turns out that packages correspond to paths in the suffix trie that can be easily stored using the suffix tree. This fact and the connection between seeds and subword complexity both contribute to the reduced linear time complexity of the seed computation algorithm.

Many other problems on seeds similar to those on covers have also been studied and are outlined below.

\subsection{Left and right seeds}\label{ssec:lrseed}
 A \df{left (right) seed} \s{u} of a string $\s{x}$ is a prefix (suffix) of $\s{x}$ that is also a seed of $\s{x}$.
 A \df{ minimal (maximal) left seed array}
of $\s{x}[1..n]$ is an integer array of length $n$  whose $i$-th element
is the minimum (maximum) length of the left seed of $\s{x}[1..i]$.
The \df{minimal (maximal) right seed array} is defined analogously. See Figure~\ref{fig:lrseedex} for an example presented in~\cite{LRSeed}.
In~\cite{Christou2013,CCIKPRRSW11} Christou {\it et al.} present a linear-time algorithm to compute the minimum and maximum left seed arrays of \s{x}. Both these algorithms rely on the linear time computation of the period and cover arrays of \s{x} (see introduction to Section~\ref{sect:cover}) in order to achieve $\bigO(n)$ running times. In the same paper, they also give an $\bigO(n^2)$ algorithm to compute all the seeds of \s{x} of length at least $k$ using the precomputed suffix, LCP, period and suffix period arrays of \s{x}.  In addition, they present an alternate $\bigO(n \log n)$-time approach to computing the shortest seeds of \s{x} that is based
on independent processing of disjoint chains in the suffix tree of \s{x}.  Further, by checking whether the shortest seed has length at least $m$, they extend this algorithm to compute the shortest seeds of length at least $m$ in $\bigO(n \log(n/m))$ time. Thus, for sufficiently large $m=\Theta(n)$, the running time of the algorithm reduces to $\bigO(n)$.
\begin{figure}[!ht]
	\begin{center}
	\footnotesize{
		\begin{tabular}{C{1cm} C{2mm} C{2mm} C{2mm} C{2mm} C{2mm} C{2mm} C{2mm} C{2mm} C{3mm} C{3mm} C{3mm} C{3mm} C{3mm} C{3mm} C{3mm}}
			& 0 & 1 & 2 & 3 & 4 & 5 & 6 & 7 & 8 & 9 & 10 & 11 & 12 & 13 & 14\\\\[-2pt]
			$\s{x}$ & a & b & a & a & b & a & b & a & a & b & a & a & b & a & b\\\\[-2pt]
			$LS_{min}$ & 1 & 2 & 2 & 3 & 3 & 3 & 3 & 3 & 3 & 3 & 3 & 3 & 3 & 3 & 3\\\\[-2pt]
			$LS_{max}$ & 0 & 0 & 2 & 3 & 4 & 5 & 6 & 7 & 8 & 9 & 10 & 11 & 12 & 13 & 14 \\\\[-2pt]
			$RS_{min}$ & 1 & 2 & 2 & 3 & 3 & 3 & 5 & 3 & 5 & 5 & 3 & 8 & 5 & 3 & 8  \\\\[-2pt]
			$RS_{max}$ & 0 & 0 & 2 & 3 & 4 & 5 & 6 & 7 & 8 & 9 & 10 & 11 & 12 & 13 & 14
		\end{tabular}
		}
	\end{center}	\vspace*{-4mm}
	\caption{$LS_{min}$, $LS_{max}$, $RS_{min}$ and $RS_{max}$ are the minimal left seed, maximal left seed, minimal right seed and maximal right seed arrays, respectively,  computed for the string $\s{x}=abaababaabaabab$ --- adapted from \cite{LRSeed}.}\label{fig:lrseedex}\vspace*{-2mm}
\end{figure}

In \cite{CCGIP12} Christou {\it et al.} describe an $\bigO(n \log n)$-time algorithm to compute the minimal right seed array. Their solution uses a variant of the partitioning algorithm introduced by Crochemore in~\cite{crochemore:81}, as employed by Iliopoulos, Moore \& Park in~\cite{imp:93}, to find the sets of ending positions of all occurrences of each distinct substring in $\s{x}$. Using this methodology, \cite{CCGIP12} finds a suffix of each prefix of the string that is covered by some substring, then checks for occurrences of right seeds to compute the minimal right seed array. They also present a simple $\bigO(n)$ algorithm to compute the maximal right seed array by detecting border-free prefixes of \s{x}.
In addition to these results, the extended journal version~\cite{LRSeed} of ~\cite{CCGIP12}
describes algorithms to compute all the left and right seeds of $\s{x}$. To compute all the left seeds of \s{x} their linear time algorithm uses the maximal cover array and the period array of \s{x}. Since the right seeds of \s{x} are just the left seeds of the reverse string $\s{x^R}$, all right seeds of \s{x} can be computed in linear time by applying the left seeds algorithm to $\s{x^R}$.

 In 2013  Flouri {\it et al.} \cite{FIKPPST13} introduce the \df{enhanced left seed}; that is, a proper prefix $\s{u}$ of $\s{x}$ that occurs at least twice in $\s{x}$ and such that the number of letters in $\s{x}$ which
lie within some occurrence of $\s{u}$ in a superstring of $\s{x}$ is a maximum over all such prefixes of $\s{x}$.  Making use of new data structures introduced in the paper, the authors describe an $\bigO(n\log n)$-time algorithm to compute the minimal (shortest length) enhanced left seed of \s{x}. The running time of the algorithm is dominated by the time required to compute some of these data structures. Then they go on to define the \df{enhanced left-seed array} --- an integer array of length $n$ whose $i$-th element is equal to the length of the enhanced left seed of the prefix of length $i$.  To compute the minimal enhanced left-seed array, they apply the minimal enhanced left seed algorithm repeatedly, thus computing the minimal enhanced left seed of every prefix of \s{x} in $\bigO(n^2)$ time.


\subsection{$\lambda$-seeds}
\label{ssec:lambdaseed}
In \cite{GZI06} Guo {\it et al.} attempted to extend the $\lambda$-covers problem of Section~\ref{ssec:kcover} to $\lambda$-seeds; that is,
given $\s{x}[1..n]$ and an integer $\lambda$, find all the sets $U = \{\s{u}_1, \s{u}_2,\ldots,\s{u}_{\lambda}\}$ of substrings of $\s{x}$ such
that:
\begin{itemize}
\item[(1)]  $|\s{u}_1| = |\s{u}_2| = \ldots = |\s{u}_{\lambda}|$;
\item[(2)] there exists a superstring $\s{y} = \s{vxw}$ of $\s{x}$ with $ |\s{v}|, |\s{w}| < |\s{u}_i|$ such that $\s{y}$ can
be constructed by concatenating or overlapping elements of $U$.
\end{itemize}

Of course the results presented in \cite{GZI06} are subject to the same difficulties raised by the counterexample of Czajka \& Radoszewski \cite{CR21}  (Section~\ref{ssec:kcover}).

\subsection{Approximate seeds}\label{ssec:approxseed}
Here again we make use of the distance measures defined in Section~\ref{ssec:approxc}.  In~\cite{CIPS05a}, Christodoulakis {\it et al.}  study the approximate seeds of strings
under the distance rules $D = H$, $D = E$ and $D = WE$ (weighted edit).

\medskip
They define a string \s{s} to be a \itbf{$t$-approximate seed} of \s{x}, $t \in \mathbb{N}$, if there exist nonempty strings $\s{s_1}, \s{s_2}, \ldots, \s{s}_r$
 such that (i) $D(\s{s}, \s{s}_i) \leq t$, for $1 \leq i \leq r$, and (ii) there exists a superstring $\s{y} = \s{uxv}$ of \s{x}, $|\s{u}| < |\s{s}|$ and $|\s{v}| < |\s{s}|$, that can be constructed by overlapping or concatenated copies of $\s{s_1}, \s{s_2}, \ldots, \s{s}_r$. They then solve the following three problems: 
\begin{itemize}
\item [(1)] \textit{Smallest distance approximate seed problem:} here strings $\s{x}[1..n]$, $\s{s}[1..m]$ and a distance function $D$ are assumed to be given and the minimum $t$ value is computed. Their solution first computes the distance between \s{s} and every substring \s{u} of \s{x} and then uses a dynamic programming approach to compute $t_i$ such that \s{s} is a $t_i$-approximate seed of $\s{x}[1..i]$. It follows that $t_n$ is the minimum $t$ such that \s{s} is a $t$-approximate seed of \s{x}. The algorithm runs in $\bigO(mn)$ time for both $D=H, E$. However for $D=WE$, it requires $\bigO(mn^2)$-time.
\item [(2)] \textit{Restricted smallest approximate seed problem:} In this case the string \s{s} is not given, and so any substring of \s{x} is a candidate for the t-approximate seed. Their solution to this problem is similar to that used to solve problem (1), but requires significantly more time --- $\bigO(n^4)$ time for $D=WE$ and $\bigO(n^3)$ time for $D=H$.
\item [(3)] \textit{Smallest approximate seed problem:} This problem is a generalization of (2) in that not only is $\s{s}$ not given, it is moreover not required to be a substring of $\s{x}$. The authors show that this problem is NP-complete for any distance rule $D$ by reduction from the NP-complete shortest common supersequence (SCS) problem~\cite{M78, ru:81}.


\end{itemize}


Further finite automaton-based algorithms for problems (1) and (2), under Hamming distance (D = H) bounded by $k$, were described in several contributions
by Guth {\it et al.}~\cite{GM10, GM09, GM09a}.
An algorithm for problem (3) was included in Guth's doctoral dissertation \cite{OG14}, which also included experimental evaluation of these algorithms.

In~\cite{KR20}, Kedzierski and Radoszewski propose efficient algorithms for computing (many) variants of approximate covers and seeds and improve
upon the complexities of previous algorithms. They show that their solutions are particularly efficient if the number (or total cost) of the allowed errors is bounded. In the context of seeds, they notably present $\bigO(n^2k)$ and $\bigO(n^3\sqrt n \log n)$ time algorithms to solve the above problem (2) for $D = H$ and $D = WE$, respectively. They also show that for $D=H$, problem (3) remains NP-hard even when the length of \s{s} is fixed, a result that holds even for strings on a binary alphabet.

\subsection{Partial seeds}\label{ssec:partialseed}
In this section we discuss the notion of partial seeds introduced in~\cite{KPRRW18a} by the same authors
(Kociumaka {\it et al.})
who introduced the partial covers~\cite{KRRPW2015} discussed in Section~\ref{ssec:alphac}.

Let $C(\s{u},\s{x})$  denote the number of positions covered by (full) occurrences of \s{u} in \s{x}. Then the non-empty prefix (suffix) of \s{x} that is also the suffix (prefix) of \s{u} is called the \df{left (right) overhanging} occurrence of \s{u} in \s{x}. $S(\s{u},\s{x})$ is the number of positions covered by the full, left and right overhanging occurrences of \s{u} in \s{x}. Then \s{u} is an \df{$\alpha$-partial seed} of \s{x}, if $S(\s{u},\s{x}) \geq \alpha$. For example, if $\s{x} = abaababaaaaba$, then $S(aba,\s{x}) = 11$ and $S(abaa,\s{x}) = 11$. Therefore both are $11$-partial seeds of \s{x}, but $aba$ is the shortest one. The authors present an $\bigO(n \log n)$ algorithm to compute all shortest $\alpha$-partial seeds of \s{x}.  Their solution uses the augmented suffix tree (cover suffix tree), originally introduced in \cite{KRRPW2015}, to compute $\alpha$-partial covers, with some additional nodes. They also describe an $\bigO(n \log n)$ time algorithm to compute a factor \s{u} of \s{x}, given an interval $[\ell, r], 0 \leq \ell \leq r \leq n$, such that $|\s{u}| \in [\ell, r]$ and which maximizes $S(\s{u},\s{x})$. Recently, by giving a linear time construction of the cover suffix tree data structure, Radoszewski in~\cite{radoszewski2023} shows that  all $\alpha$-partial seeds can be computed in $\bigO(n)$-time.


\subsection{Extensions to indeterminate and weighted strings}\label{ssec:indetseeds}

In Section~\ref{ssec:coverindet} we discussed $q$-conservative indeterminate strings, and the
problem of finding covers in such strings. Here we mention the $\lambda$-conservative seeds problem introduced by Antoniou {\it et. al} in~\cite{ACIJL08}: given a $q$-conservative indeterminate string \s{x} and $\lambda \in \mathbb{Z^+}$, a $\lambda$-conservative seed is a seed of \s{x} of length $\lambda$. Making use of the Aho-Corasick automaton~\cite{AC75}, the authors describe an $\bigO(n\lambda)$-time algorithm to compute the $\lambda$-conservative seeds (if they exist) of \s{x}.

Section~\ref{ssec:coverindet} also discussed weighted strings and cover algorithms for them. In \cite{ZGI10} Zhang {\it et al.} discuss \textit{loose} and \textit{strict} matching (see Section~\ref{ssec:coverindet}) and present two cover algorithms for weighted strings based on these string matching variations. In the same paper, they also propose $\bigO(n^2)$-time algorithms that compute all seeds in the given weighted string based on these matchings.

\section{Open problems}
\label{sect-open}
As suggested in the Introduction, a central motivation for the study of covers and seeds is the ubiquitous requirement to find compressed representations of long strings that moreover disclose patterns --- some sort of ``meaning'' --- not evident in their original linear formulation.
In this context we present here a collection of
open problems arising from the work surveyed
above, some suggested by the authors themselves, some of them new.
\subsection{Covers}
\label{sectopen-1}
\begin{description}

\item[Find String on Minimum Alphabet
(\ref{sect:cover}):]
\cite{GKRRW20} describes a linear-time approach to
determining a string corresponding to a given border array
or prefix array, but not necessarily on a minimum alphabet.
Can this improvement be achieved, also in linear time?

\item[Shortest Covers, All Rotations
(\ref{ssec:coverca}):]
\begin{itemize}
\item[(1)] The question of the space required for shortest cover computation has also been raised.  In \cite{GRS19} two space-efficient near-linear randomized algorithms are described that with high accuracy compute the shortest cover of given \s{x}.  Can the shortest cover be computed in polylog$(n)$ space?

\item[(2)] \cite{CCRR20-1,CCRR21-1,CCRR20-2} all employ $\ST_{\s{x}^3}$ to represent the seeds of \s{x}.  Can a more direct approach be found, possibly replacing the suffix tree, possibly reducing processing time?
\end{itemize}

\item[$k$-Covers (\ref{ssec:kcover}):]
In view of the NP-completeness of the original $k$-cover problem, and the difficulty discovered in \cite{CR21} with its replacement, further results in this challenging area would be very welcome.

\item[$\lambda$-Covers (\ref{ssec:kcover}):]
\begin{itemize}
    \item[(1)] The complex work of Radoszewski \& Straszy\'{n}ski \cite{RS20} deals efficiently with the case $\lambda = 2$. Their solution also deals with the generalized case of $\lambda > 2$; however, for each unit increase in $\lambda$, the running time complexity of the solution increases by a factor of $n$. Does a solution with better running time for $\lambda > 2$ exist?
 \item[(2)] In \cite{RS20} the authors propose the following problems:
 \begin{itemize}
 \item[(a)] Can a shortest $2$-cover be computed in linear time?
 \item[(b)] Can we efficiently compute a variation of the $2$-covers (and $\lambda$-covers) problem, in which the factors that cover the string are of different lengths?
\end{itemize}
\end{itemize}


\item[Maximal Covers (\ref{ssec:oc}):]

\begin{itemize}
\item [(1)] As noted in Section~\ref{ssec:alphac}, the all-partial-covers can be computed in $\bigO(n)$-time using the linear time construction of the cover suffix tree data structure. Can we compute the all-partial-covers/maximal covers without the need to use annotated suffix trees?
\item [(2)] Are the maximal covers of practical use for the compact representation of any classes of string?  
                If so, can the computation of this representation be iterated?
\end{itemize}
\item[ACP Problems (\ref{ssec:approxc}):]
In \cite{ALLP19,ALLLP19} the authors propose related problems:
\begin{itemize}
    \item [(1)] Does ACP remain NP-hard on a constant alphabet?
    \item [(2)] What is the effect on ACP complexity if distance metrics other than Hamming are used?
    \item [(3)] In addition to RACP, are there other
    relaxations of ACP solvable in polynomial time?
\end{itemize}
\item[Specialized Covers (\ref{ssec:specialc}):]
In \cite{MIBT14} the authors ask whether all the Abelian covers of given \s{x} can be computed in less than $\bigO(n^2)$ time.  The question has to some degree been answered affirmatively by Kociumaka {\it et al.} \cite{KRW17}, who describe an $\bigO(n^2/\log n)$ algorithm for this and other related problems, given a constant-sized alphabet.  However, in view of the recent proof \cite{RRSWZ21a} that the closely related Abelian squares problem is ``3SUM'' hard, it seems unlikely that a clearly more efficient algorithm, free of the $n^2$ factor, can be found.
Also, the following questions arise:
\begin{itemize}
\item [(1)] How often can strings on a small alphabet be covered by an Abelian cover?
\item [(2)] Can these ``coverable'' strings be characterized in a useful way?
\end{itemize}
Similarly, how many strings possess a cyclic cover --- significantly more than those with just a cover?
\end{description}
\subsection{Seeds}
\label{sectopen-2}
\begin{description}
\item[Left and Right Seeds (\ref{ssec:lrseed}):]
In~\cite{Christou2013, CCGIP12} the authors pose the following problem: given an integer array $A$ of length $n$, determine whether or not $A$ is the minimal left-seed (resp. right-seed) array of some string and, if so, construct one such string. In addition, the following questions are of interest:
\begin{itemize}
    \item [(1)] In the above problem, what would be the minimum number of distinct letters required to build such a string? Can such strings always be constructed over a bounded alphabet?
    \item [(2)] Can we compute the  minimal right seed array in linear time?
\end{itemize}

\item[Enhanced Left Seeds and Enhanced Left Seeds Array (\ref{ssec:lrseed}):]
As noted in \cite{FIKPPST13}, the same problem arises here as with the
Abelian covers calculation: can the minimal enhanced left-seed array be computed in time clearly less than order $n^2$?  Moreover:
\begin{itemize}
\item [(1)] Does an $\bigO(n \log n)$ or $\bigO(n)$ algorithm exist to compute this array?
\item [(2)] What is the complexity of an optimal algorithm to compute the \df{maximal} enhanced left-seed array? Does an $\bigO(n \log n)$ or $\bigO(n)$ algorithm exist for this problem?
\end{itemize}

\item[$\lambda$-Seeds (\ref{ssec:lambdaseed}):]
As with $\lambda$-covers, can the $\lambda$-seeds problem be usefully reformulated?

\item[Approximate Seeds (\ref{ssec:approxseed}):]


Do finite automata-based solutions exist for problems (1) and (2) under $D = E, WE$, and bounded by a constant $k$? If so what are the time complexities of these solutions?
\item [Partial Seeds (\ref{ssec:partialseed}):]
Can we compute partial seeds using the $\SA$ and $\LCP$ arrays? If so, can they also be computed in $\bigO(n)$ time?
\end{description}

\section{Conclusion}
In this paper we have attempted to bring together in an organized fashion all the results related to covers/seeds published since the invention of these concepts more than 30 years ago. That they have been so much studied testifies to their current relevance as well as to their potential future impact on the development of combinatorics on words and string algorithms. We anticipate that further study of the numerous open problems stated here will lead to significant advances in these fundamental computational areas.

\subsection*{Acknowledgements}

\itbf{Revisions}: The authors wish to acknowledge the fine work of several referees whose insightful and knowledgeable commentary has resulted in significant improvement to this paper.

\smallskip
\itbf{Funding}: The second author was funded by the Natural Sciences \& Engineering Research Council of Canada [Grant Number 10536797].

\smallskip
\itbf{Conflict of Interest:} Both authors declare that they have no conflict of interest.

\smallskip
\itbf{Informed Consent:} This article does not contain any studies on human participants or animals performed by any of the authors.

\end{document}